\documentclass[draft]{aipproc}

\layoutstyle{6x9}

\begin{document}

\title{Burst Statistics Using the Lag-Luminosity Relationship}

\author{D. L. Band}
{address={Code 661, NASA/Goddard Space Flight Center,
Greenbelt, MD 20771} }

\author{J. P. Norris}
{address={Code 661, NASA/Goddard Space Flight Center,
Greenbelt, MD  20771} }

\author{J. T. Bonnell}
{address={Code 661, NASA/Goddard Space Flight Center,
Greenbelt, MD  20771} }

\begin{abstract}
Using the lag-luminosity relation and various BATSE
catalogs we create a large catalog of burst redshifts, peak
luminosities and emitted energies. These catalogs permit us
to evaluate the lag-luminosity relation, and to study the
burst energy distribution.  We find that this distribution
can be described as a power law with an index of
$\alpha=1.76\pm 0.05$ (95\% confidence), close to the
$\alpha=2$ predicted by the original quasi-universal jet
model.
\end{abstract}

\maketitle

\section{Introduction}

Jet models predict the distribution of the
isotropic-equivalent energy $E_{iso}$:  quasi-universal jet
profile models predict an approximate power law
distribution with index $\alpha=2$, where $N(E_{iso})
\propto E_{iso}^{-\alpha}$\cite{1}. The
isotropic-equivalent energy $E_{iso}$ is the total energy
radiated if the observed flux were radiated isotropically.
To study the distribution of burst intensities, we used the
lag-luminosity relationship to create a burst database with
redshifts, peak luminosities, and burst energies, and then
we fit energy distributions to the burst database.  Of
course, this database can be used for other studies.

In the lag-luminosity relation\cite{2} the peak bolometric
luminosity $L_B$ is a function of the lag $\tau_B$ between
two energy bands in the burst's frame---$L_B=Q(\tau_B)$.
But $\tau_0$ is measured in our frame. We model
$\tau_B=(1+z)^c \tau_0$:  time dilation contributes -1 to
c, while the redshifting of temporal structure with a
smaller lag from higher energy contributes $\sim$1/3
(pulses are narrower at high energy). The peak bolometric
luminosity is related to the peak bolometric energy flux
$F_B = L_B / [4\pi D_L^2]$, where $D_L$ is the luminosity
distance. The peak bolometric energy flux is related to the
peak photon flux P (integrated over an energy band, e.g.,
50--300~keV for BATSE data): $F_B = \langle E\rangle P$.
The result is an implicit equation that must be solved for
each burst:
\begin{equation}
P = Q\left((1+z)^c \tau_0 \right) / \left[\langle E \rangle
   4\pi D_L^2\right]
\end{equation}
After solving eq.~1 for the redshift, $L_B$ and $E_{iso}$
can be calculated from $F_B$ and the energy fluence,
respectively.

The original lag-luminosity relation was a single power
law, e.g., $L_B\propto \tau_B^{-1.15}$.  But this power law
over-predicts the luminosity of GRB980425 (assuming this
burst was SN1998bw). Consequently Salmonson\cite{3} and
Norris\cite{4} suggested breaking the single power law; for
$\tau_B>$0.35~s the power law index is -4.7. A population
of nearby, long lag bursts resulted.

With a database of bursts with $E_{iso}$ we can now
calculate the energy distribution.  The methodology
presented here\cite{5} can also be applied to the
luminosity function.  The probability of detecting a given
energy is truncated by the detection threshold: $p(E_{iso}
\,|\, E_{iso,th}M(\vec{a}))$ where $E_{iso,th}$ is the
threshold value of $E_{iso}$ for that burst and
$M(\vec{a})$ is the model (e.g., the functional form of the
energy distribution) with parameters $\vec{a}$.  For the
ensemble of bursts the probability of detecting bursts with
the observed energies is
\begin{equation}
\Lambda = \prod_i p\left(E_{iso,i} \,|\, E_{iso,th,i}
   M(\vec{a}) \right)
\end{equation}
where the product is over each burst.  This probability is
the likelihood for the model $M(\vec{a})$.  In frequentist
statistics, we maximize $\Lambda$ with respect to the
parameters $\vec{a}$ to get a best fit value. In Bayesian
statistics the likelihood is a factor in the ``posterior,''
which can be used for confidence ranges and best fit
values; the Bayesian approach allows the use of ``priors''
reflecting our expectations for $\vec{a}$.

Note that to study the energy distribution we do NOT need a
complete sample in terms of observed fluences, only a
sample that has no bias on the intrinsic $E_{iso}$. There
can be gaps in the distribution of peak fluxes, but
$E_{iso}$ has to be drawn uniformly from $p(E_{iso} \,|\,
E_{iso,th}M(\vec{a}))$ in our sample. On the other hand, if
we want the burst rate per comoving volume as a function of
redshift, then we do need a complete sample.

But is the resulting energy distribution a good
representation of the data? The likelihood (frequentist
approach) or posterior (Bayesian approach) can be used to
compare models (functional forms), but do not indicate
``goodness-of-fit.''  However, our methodology assumes the
energies are drawn uniformly from $p(E_{iso} \,|\,
E_{iso,th}M(\vec{a}))$. The cumulative distribution of
$p(E_{iso} \,|\, E_{iso,th}M(\vec{a}))$ should therefore be
a straight line, and the average value should be 1/2, with
a statistical uncertainty of $[12N]^{-1/2}$ for $N$ bursts.

\section{Results}

We started with 1438 BATSE bursts for which we calculated
lags. Of these, 1218 have positive lags.  These bursts also
have hardness ratios, peak fluxes and durations.  To
calculate the average energy $\langle E\rangle$ we used the
``GRB'' spectral fits of Mallozzi et al.\cite{6} to the
peaks of 580 of these bursts.  For the 858 bursts without
fits we assumed average spectral indices $\alpha=-0.8$ and
$\beta=-2.3$. Plotting HR$_{32}$ (the 100--300~keV to
50--100~keV hardness ratio) vs. $E_p$ shows a clear
correlation which can be approximated by $E_p$=240
HR$_{32}^2$~keV; we used this relation for the bursts
without spectral fits.

Redshifts were calculated for this database for both the
original simple power law lag-luminosity relation and the
broken power law Salmonson\cite{3} and Norris\cite{4}
introduced to incorporate GRB980425. As expected, the
difference in the lag-luminosity relations is apparent at
low redshifts: the broken power law results in a population
of nearby bursts. There were few physically implausible
high $z$ bursts (e.g., $z>20$) and thus no additional
cutoffs on the lag-luminosity relation are required. In the
absence of additional information, the choice between the
two lag-luminosity relations depends on whether GRB980425
is considered to be a typical low luminosity burst.  For
the remainder of this analysis we use a single power law
lag-luminosity relation.

As an aside, we found that the redshift calculation is
sensitive to the value of $\langle E\rangle$. Calculating
this quantity inconsistently can introduce errors into the
resulting database.

We calculated the energy $E_{iso}$ for each burst from the
redshift and energy fluence. The results were reasonable
(see Fig.~1): few bursts had $E_{iso}>10^{54}$~erg or
$E_{iso} < 10^{51}$~erg.

\begin{figure}
  \includegraphics[height=.3\textheight]{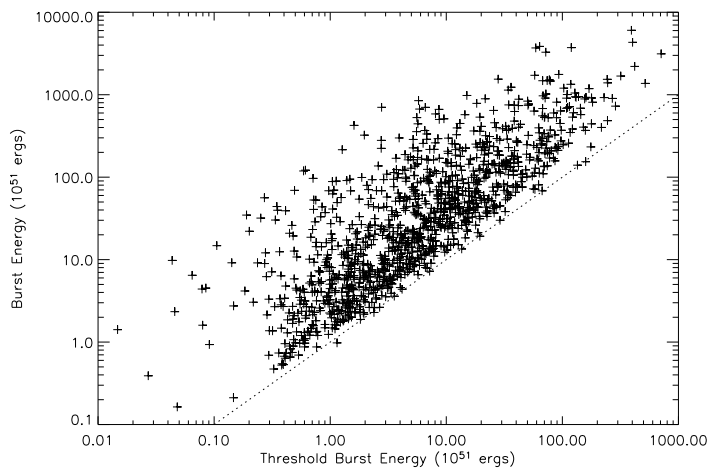}
  \caption{Scatter plot of the isotropic equivalent
energy $E_{iso}$ vs. the detection threshold.}
\end{figure}

The energy detection threshold $E_{iso,th}$ can be
calculated by scaling $E_{iso}$ by the ratio of the
threshold peak photon flux to the observed peak flux.
BATSE's threshold peak flux was $P_{min}\sim$0.3 ph
cm$^{-2}$ s$^{-1}$; however, the number of bursts in our
sample with $E_{iso}$ just above the threshold is
suspiciously low (see Fig.~1), suggesting that the sample's
true threshold was greater than 0.3 ph cm$^{-2}$ s$^{-1}$.
Consequently we used $P_{min}\sim$0.5 ph cm$^{-2}$ s$^{-1}$
as the threshold, deleting bursts with $P<0.5$ ph cm$^{-2}$
s$^{-1}$.

The left hand side of Fig.~2 shows the likelihood surface
for our sample assuming a power law functional form, where
the two parameters are the low energy cutoff $E_2$ and the
power law index $\alpha$ (i.e., $N(E_{iso})\propto
E_{iso}^{-\alpha}$ for $E_{iso}\ge E_2$). The likelihood is
maximized by $E_2$ equal to the lowest observed value
$E_{iso}$, although lower values are not ruled out.  The
best fit spectral index is $\alpha=1.76\pm0.05$ (95\%
confidence). Although $\langle P(>E_{iso}) \rangle$=
0.4642$\pm $0.0089 ($N$=1054, assuming $P_{min}\sim$0.5 ph
cm$^{-2}$ s$^{-1}$) deviates from 1/2 by $4\sigma$,
considering the possible systematic errors (e.g., in the
estimation of $E_p$ from the hardness ratio), this value of
$\langle P(>E_{iso})\rangle$ indicates that a power law
energy distribution is a fairly good characterization of
the data.

\begin{figure}
  \includegraphics[height=.25\textheight]{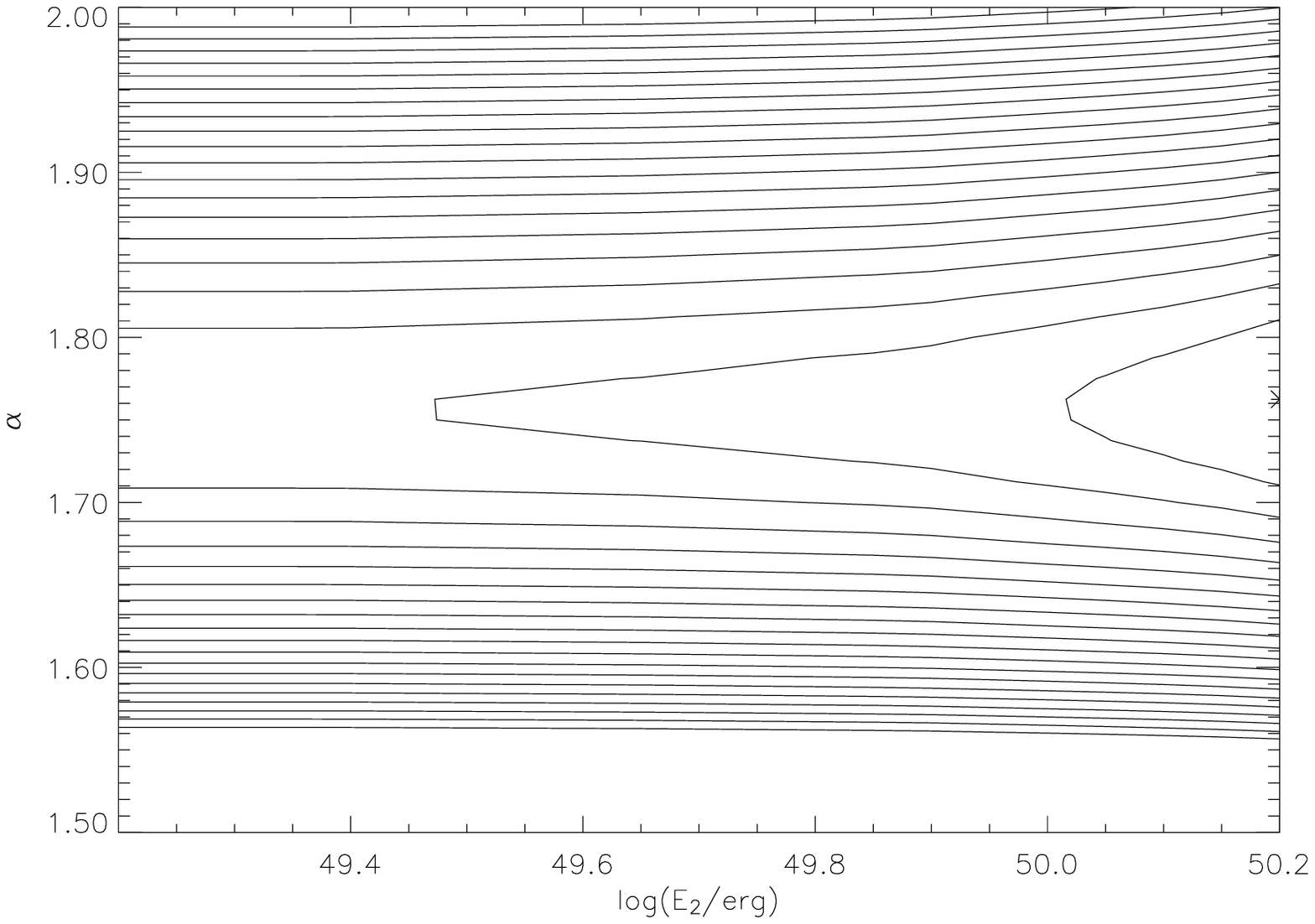}
  \includegraphics[height=.25\textheight]{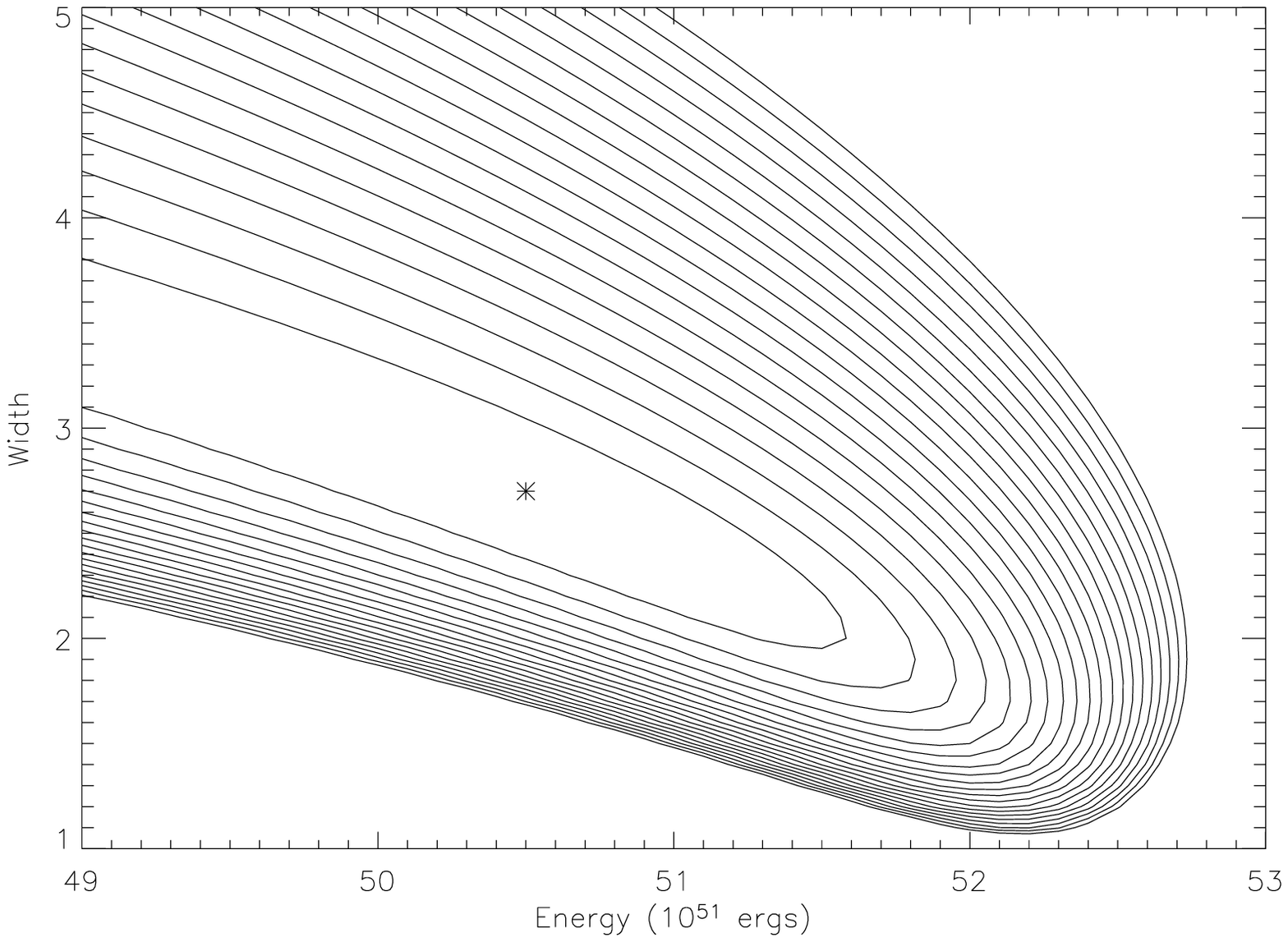}
  \caption{Contour plots of the likelihood surface for a
power law energy distribution (left) and lognormal energy
distribution (right).  The power law has a low energy
cutoff $E_2$ and power law index $\alpha$; the contours are
spaced by $\Delta$log(likelihood)=1. The lognormal
distribution has a central value $E_{\rm iso,cen}$ and a
logarithmic width $\sigma_E$; the contours are spaced by
$\Delta$log(likelihood)=10.}
\end{figure}

We also tried a lognormal energy distribution (right hand
side of Fig.~2). The maximum likelihood occurs at $E_{\rm
iso,cen} = 3 \times 10^{51}$~ergs and $\sigma_E = 2.7$. The
surface's shape indicates that the data permit a high
central value of $E_{iso}$ and a narrow distribution, or a
low central value of $E_{iso}$ and a broad distribution.
The observational cutoff truncates the true energy
distribution, and the low energy extent is relatively
unknown. We find $\langle P(>E)\rangle$=0.4821$\pm$0.0089
($N$=1054, assuming $P_{min}\sim$0.5 ph cm$^{-2}$
s$^{-1}$), consistent with 1/2 at the 2$\sigma$ level.

\section{Implications}

A quasi-universal jet profile that is a power law in the
off-axis angle $\theta$---the energy per solid angle
$\epsilon(\theta)\propto\theta^{k}$---results in a power
law energy distribution (or luminosity function) with index
$\alpha=1-2/k$ (hence $\alpha=2$ for $k=-2$), while a
Gaussian profile results in $\alpha=1$.  Lloyd-Ronning et
al.\cite{1} found that if the profile parameters are
distributions, the luminosity functions could be
approximated by power laws with $\alpha\sim 2$ for power
law profiles and $\alpha\sim 1$ for Gaussian profiles, but
with curvature.  The additional degrees of freedom
introduced by varying the parameters give the jet models
the freedom to fit a wide variety of energy distribution
shapes.  We find that our burst data can be fit by a power
law energy distribution with $\alpha=1.76\pm0.05$ (95\%
confidence); considering only the statistical uncertainty
the power law distribution is formally not a good fit, but
with the likely systematic uncertainties the power law
distribution is probably a good description of the data.
While our power law fit is inconsistent with the original
jet profile model ($k=-2$ and therefore $\alpha=2$), it is
consistent with the jet profile models where parameters are
permitted to vary.

A log-normal energy distribution also describes the data;
the data permit a smaller average energy if the
distribution is wider.

\end{document}